\documentclass[aps,twocolumn,floatfix,groupedaddress,nofootinbib]{revtex4}
\usepackage{csquotes}
\usepackage{graphicx,color}
\usepackage{float}
\usepackage[caption=false]{subfig}
\usepackage{amsmath}
\usepackage{amsmath}
\usepackage{multirow}
\usepackage{dsfont}

\begin{document}

\title{Structure and thermodynamics of two dimensional Yukawa liquids}
\author{F. Lucco Castello and P. Tolias}
\affiliation{Space and Plasma Physics, Royal Institute of Technology, Stockholm, SE-100 44, Sweden}
\begin{abstract}
\noindent The thermodynamic and structural properties of two dimensional dense Yukawa liquids are studied with molecular dynamics simulations. The "exact" thermodynamic properties are simultaneously employed in an advanced scheme for the determination of an equation of state that shows an unprecedented level of accuracy for the internal energy, pressure and isothermal compressibility. The "exact" structural properties are utilized to formulate a novel empirical correction to the hypernetted-chain approach that leads to a very high accuracy level in terms of static correlations and thermodynamics.
\end{abstract}
\maketitle

\section{Introduction} \label{intro}

\noindent The two-dimensional Yukawa one-component plasma (2D-YOCP) consists of charged point particles that are confined on a two-dimensional surface and are immersed in a polarizable neutralizing background with their interactions described by the Yukawa (screened Coulomb) pair potential $\phi(r)=(Q^2/r)\exp\left(-r/\lambda\right)$. Here $Q$ is the particle charge and $\lambda$ the screening length defined by the polarizable background. Thermodynamic state points of the 2D-YOCP are specified by two dimensionless variables\,\cite{fortov2005,morfill2009}: the coupling parameter $\Gamma=\beta Q^2/d$ and the screening parameter $\kappa=\lambda/d$, where $\beta=1/(k_{\mathrm{B}}T)$ with $k_{\mathrm{B}}$ Boltzmann's constant and where $d=(\pi n)^{-1/2}$ is the 2D Wigner-Seitz radius with $n$ the particle density. The coupling parameter provides a measure of the strength of the unscreened particle interactions with the strong coupled (liquid) regime characterized by $\Gamma\gtrsim1$\,\cite{donko2008,bonitz2010}, while the screening parameter dictates the interaction softness that varies from infinitely long-ranged Coulomb-like in the one-component plasma (2D-OCP) limit of $\kappa\to0$ to extremely short-ranged hard sphere-like in the opposite limit of $\kappa\to\infty$.

The 2D-OCP system has been long known to be relevant for classical electron layers trapped over the surface of liquid helium\,\cite{grimes1979,fisher1979,totsuji1980,baus1980}. In more recent years, a renewed interest in 2D-YOCP systems was sparked by the observation that they can adequately model dust monolayers levitating in the sheath region of low temperature noble gas discharges\,\cite{konopka2000,ratynskaia2005,ratynskaia2006,nosenko2006,feng2010} and to investigate combustion phenomena in confined geometries\,\cite{yurchenko2017}. As a consequence, a considerable amount of effort has been dedicated to investigate the phase behavior\,\cite{hartmann2005,klumov2010,hartmann2010,yazdi2015} as well as the structural\,\cite{murillo2003,vaulina2006,ott2015}, thermodynamic\,\cite{totsuji2004,feng2016,khrapak2016,kryuchkov2017,huang2017,li2017} and dynamic properties\,\cite{golden2000,ott2009,hartmann2011,ott2014,khrapak2018} of the 2D-YOCP.

This work focuses on two issues which are still not fully resolved for 2D-YOCP liquids: (a) the acquisition of an accurate equation of state through the reduced excess internal energy that can also be employed to accurately estimate other thermodynamic properties over the entire range of screening parameters relevant to experimental realizations of Yukawa systems, (b) the development of an accurate integral equation theory approach that would allow for the reliable computation of structural properties without necessarily resorting to computer simulations.

In order to address the first issue, systematic molecular dynamics simulations are carried out in the entire 2D-YOCP liquid regime that are utilized for direct extraction of the internal energy, pressure and inverse isothermal compressibility. A novel approach is then presented that simultaneously utilizes these exact thermodynamic data in order to acquire an equation of state through the internal energy that is robust with respect to thermodynamic integration and thermodynamic differentiation.

In order to address the second issue, systematic long molecular dynamics simulations are performed in the entire 2D-YOCP liquid regime that are employed for the direct extraction of the radial distribution function and its characteristic functional features. In the absence of a straightforward way to adapt to the 2D-YOCP advanced ultra-accurate integral equation theory approaches that are available for the 3D-YOCP\,\cite{lucco2021c}, the exact data for the magnitude of the global radial distribution function maximum are used in order to construct an empirical modification to the hypernetted-chain approach. In spite of its simplicity, the emerging approximation leads to very accurate predictions for the structural (and thermodynamic) properties of 2D-YOCP liquids.

The paper is organized as follows. In Section \ref{sec:md}, the  molecular dynamics simulations are presented and their results are discussed. In Section \ref{sec:eos}, the simulation data are employed for the determination of a new 2D-YOCP liquid equation of state which is extensively compared to other equations of state that are already available in the literature. In Section \ref{sec:iet}, the simulation data are employed for the construction of a novel integral equation theory approximation of empirical nature, whose accuracy and validity region are quantified. In Section \ref{sec:summary}, the results are summarized and possible future developments are discussed.

\section{Molecular dynamics simulations}\label{sec:md}

\subsection{Simulation parameters}\label{sec:md_parameters}

\begin{figure}[t]
	\centering
	\includegraphics[width=3.1in]{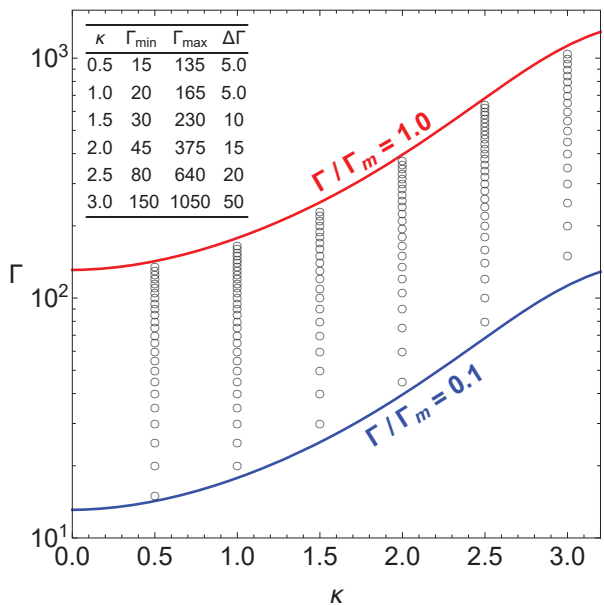}
	\caption{The 2D-YOCP state points that are investigated with NVT MD simulations in the $(\log\Gamma-\kappa)$ phase diagram. All the simulated state points (black open circles) belong to the dense fluid region of the phase diagram which roughly extends between the curve $\Gamma=0.1\Gamma_{\mathrm{m}}(\kappa)$ (blue) and the melting curve $\Gamma=\Gamma_{\mathrm{m}}(\kappa)$ (red). Here $\Gamma_{\mathrm{m}}(\kappa)$ denotes the coupling parameter resulting from the analytical parametrization of the melting line of Eq.\eqref{eq:Gamma_melting}\,\cite{hartmann2005}. The exact numerical values for the simulated state points can be retrieved from the information provided in the inset table: for each screening parameter, $\kappa$, the coupling parameter was augmented with a constant step $\Delta\Gamma$ between $\Gamma_{\mathrm{min}}$ and $\Gamma_{\mathrm{max}}$.}\label{fig:phase_diagram}
\end{figure}

\noindent Molecular dynamics (MD) simulations were performed to determine the thermodynamic and structural properties of dense 2D-YOCP liquids. MD simulations were carried out for nearly $150$ state points characterized by screening parameters that are relevant to experimental 2D-YOCP realizations\,\cite{nosenko2006,feng2010,hartmann2010}, \emph{i.e.} $\kappa=\{0.5,\,1,\,1.5,\,2,\,2.5,\,3\}$, and coupling parameters that cover the entire dense fluid portion of the phase diagram, \emph{i.e.} $0.1\leq\Gamma/\Gamma_{\mathrm{m}}(\kappa)\leq1.0$. The state points have been summarized in Fig. \ref{fig:phase_diagram}. In the above, $\Gamma_{\mathrm{m}}(\kappa)$ is the 2D YOCP melting line as determined by the analytical parametrization\,\cite{hartmann2005}
\begin{equation}
\Gamma_{\mathrm{m}}(\kappa) = \frac{\Gamma_{\mathrm{m}}^{\mathrm{OCP}}}{1-0.388\kappa^2 + 0.138\kappa^3 - 0.0138\kappa^4}\,,\label{eq:Gamma_melting}
\end{equation}
where $\Gamma_{\mathrm{m}}^{\mathrm{OCP}}=131.0$ denotes an approximation for the 2D-OCP melting point that is consistent with the results of both computer simulations\,\cite{gann1979} and experiments\,\cite{grimes1979}.

All the simulations were performed with the LAMMPS package\,\cite{plimpton1995} and employed 4096 particles in the canonical NVT ensemble. The dynamics was resolved with a time-step of $\Delta\tau=0.001d\sqrt{\beta m}$ while the interaction potential was truncated at $r=20d$ for $\kappa=0.5$ and at $r=10d$ for $\kappa>0.5$. Two set of simulations were performed: one set of \emph{short MD simulations} consisting of $2^{19}$ time-steps for equilibration followed by $2^{19}$ time-steps for statistics (that were employed to collect 2048 samples for the thermodynamic properties) and one set of \emph{long MD simulations} consisting of $2^{19}$ time-steps for equilibration followed by $2^{24}$ time-steps for statistics (that were employed to collect 65536 samples for the structural properties).

\subsection{Structural properties}\label{sec:md_structural}

\begin{figure}[t]
	\centering
	\includegraphics[width=3.40in]{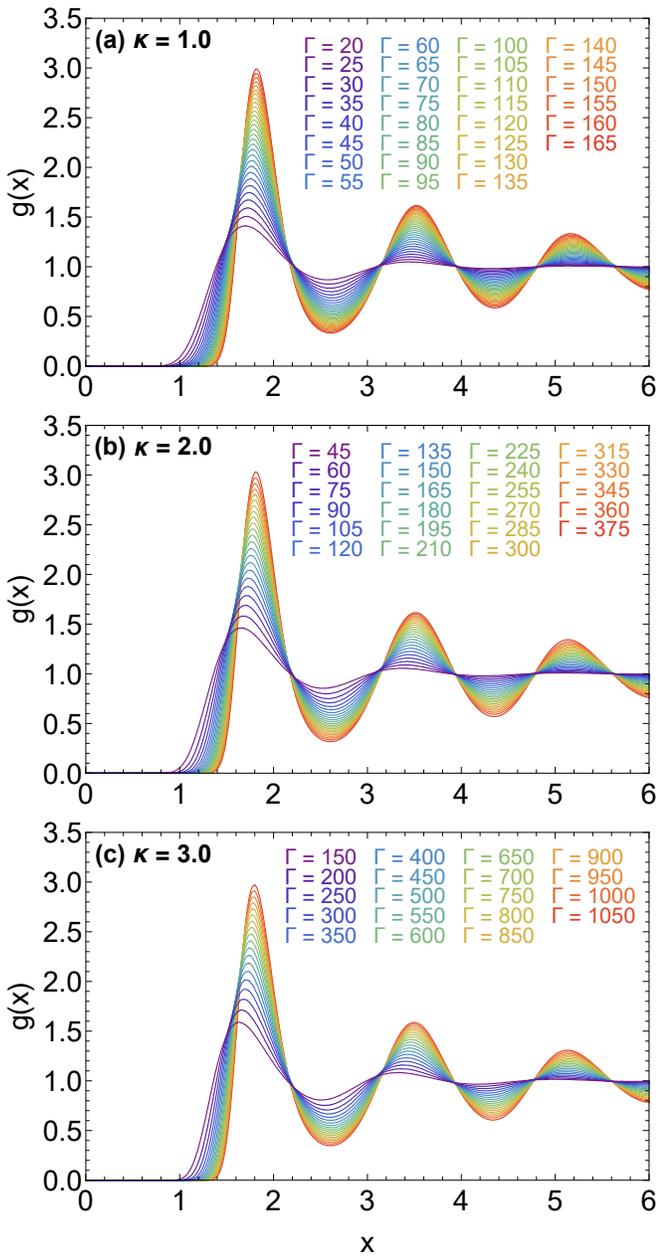}
	\caption{Radial distribution functions extracted from \emph{long MD simulations} with the histogram method. Results for constant screening parameter [$\kappa=1$ in (a), $\kappa=2$ in (b), $\kappa=3$ in (c)] and varying coupling parameters within $0.1\leq\Gamma/\Gamma_{\mathrm{m}}(\kappa)\leq1$.}\label{fig:rdf_MD}
\end{figure}

\noindent Concerning structural properties, the focus lied on the radial distribution function, $g(r)$, which was extracted from the \emph{long MD simulations} with the histogram method\,\cite{allen1989} for a bin-width of $\Delta r=0.002d$. The narrow bin-width was selected so that the magnitude of the first $g(r)$ maximum is determined with very high accuracy, since it constitutes the MD simulation input that is employed for the construction of our integral equation theory approximation, see section \ref{sec:iet} for details. Naturally, such a narrow bin-width necessitates longer simulations, since a large number of uncorrelated samples are necessary to obtain highly-resolved radial distribution functions which are unaffected by the omnipresent statistical noise. Some examples of the radial distribution functions obtained from the long MD simulations are illustrated in Fig.\ref{fig:rdf_MD}, where it is apparent that the $g(r)$ curves are subject to negligible statistical errors. Further support for the accuracy of the present radial distribution functions comes from the observation that some key figures of merit including the magnitude and position of the first maximum, first non-zero minimum and second maximum are consistent with results available in the literature from Langevin Dynamics simulations for $\kappa=\{0.5, 1.0\}$\,\cite{ott2015}. Tabulated values of the basic $g(r)$ figures of merit are provided in the supplementary material\,\cite{supplementary} for all the state points summarized in Fig.\ref{fig:phase_diagram}.

\subsection{Thermodynamic properties}\label{sec:md_thermo}

\noindent Concerning thermodynamic properties, the focus lied on the internal energy ($U$), pressure ($P$) and inverse isothermal compressibility [$K_{T}=-V(\partial P/\partial V)_T$] with $V=N/n$ the volume of a homogeneous system with $N$ particles and $n$ density. In two dimensions, $V$ is strictly the area. However, aiming to be consistent with the nomenclature developed for three-dimensional systems, we shall still refer to it as volume also for two-dimensional systems. In what follows, we shall discuss normalized (reduced) thermodynamic properties with the internal energy expressed as $u=U/(Nk_{\mathrm{B}}T)$, the pressure as $p=P/(nk_{\mathrm{B}}T)$ and the inverse isothermal compressibility as $\mu=K_T/(nk_{\mathrm{B}}T)$. In addition, since the ideal gas contributions are all known, we shall ignore them and discuss exclusively the reduced excess thermodynamic properties that are emerging from the interaction part of the Hamiltonian.

The reduced excess internal energy is related to the ensemble averaged total potential energy per particle, $u_{\mathrm{ex}}=\langle\mathcal{U}\rangle/N$ with $\mathcal{U}=\sum_i^N\sum_{j>i}^N\beta\phi(r_{ij})$ and $r_{ij}$ the distance between any particle pair\,\cite{hansen2006}. The reduced excess pressure is related to the (two-dimensional) microscopic virial $\mathcal{W}=-(1/2)\sum_i^N\sum_{j>i}^N\beta w(r_{ij})$ with $w(r)=r[d\phi(r)/dr]$, through the virial equation $p_{\mathrm{ex}}=\langle\mathcal{W}\rangle/(NV)$\,\cite{hansen2006}. The reduced excess inverse isothermal compressibility is obtained from the so-called hypervirial theorem from which it follows that $\mu_{\mathrm{ex}}=[\langle\mathcal{W}\rangle-\langle(\delta\mathcal{W})^2\rangle+\langle\mathcal{X}\rangle]/N$, with $(\delta\mathcal{W})^2=\mathcal{W}^2-\langle\mathcal{W}\rangle^2$ the microscopic virial fluctuations, $\mathcal{X}=(1/4)\sum_i^N\sum_{j>i}^N\beta x(r_{ij})$ the (two-dimensional) microscopic hypervirial and $x(r)=r[dw(r)/dr]$\,\cite{allen1989}.

Therefore, in order to obtain $u_{\mathrm{ex}}$, $p_{\mathrm{ex}}$, $\mu_{\mathrm{ex}}$ from MD simulations, it is sufficient to collect samples for $\mathcal{U}$, $\mathcal{W}$, $\mathcal{X}$ at regular time-intervals in the course of the simulation, to invoke the ergodic hypothesis for the computation of the ensemble averages and to utilize the aforementioned expressions. This procedure was followed for the \emph{short MD simulations}, since it was observed that relatively few samples are required for the determination of $u_{\mathrm{ex}}$, $p_{\mathrm{ex}}$, $\mu_{\mathrm{ex}}$ with a negligible statistical uncertainty.

\begin{figure*}[htbp]
	\centering
	\includegraphics[width=7.0in]{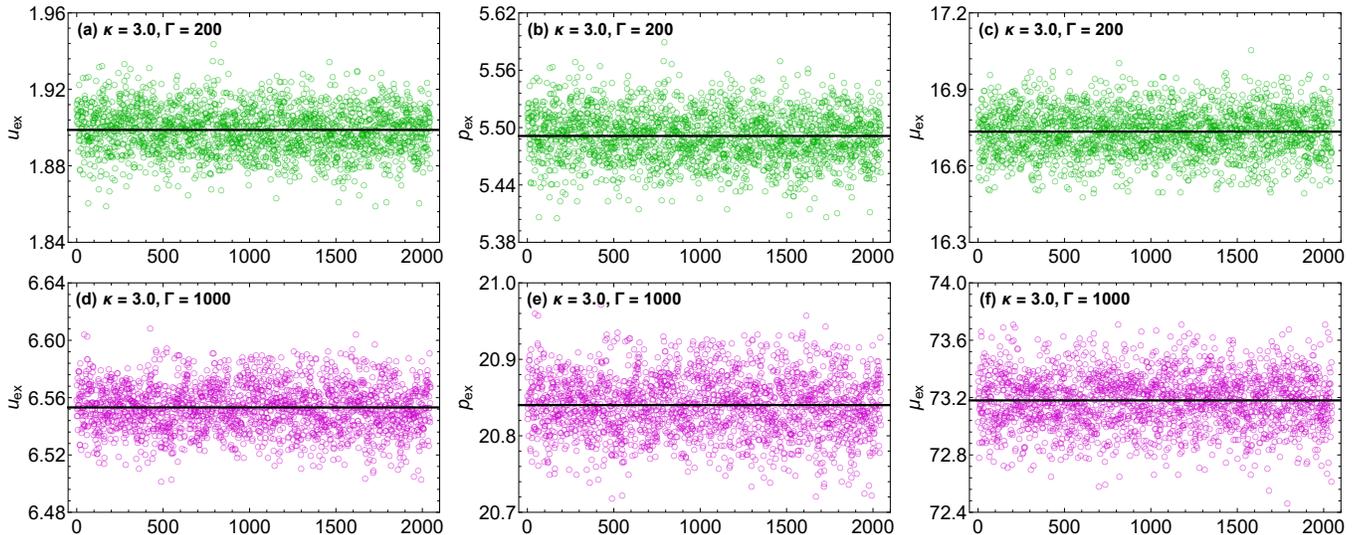}
	\caption{Samples collected for the excess internal energy [panels (a), (d)], excess pressure [panels (b), (e)] and excess inverse isothermal compressibility [panels (c), (f)] from the \emph{short MD simulations} at two 2D-YOCP state points that are defined by $\kappa = 3.0$ and $\Gamma=200$ (green), $\Gamma=1000$ (magenta). In each plot, the sample average has been demarcated with a black line.}\label{fig:MD_fluctuations}
\end{figure*}

Fig.\ref{fig:MD_fluctuations} illustrates two sample collection examples for all three thermodynamic properties at two 2D-YOCP state points characterized by $\kappa=3.0$. It is evident that all samples fluctuate around a constant average value without systematic deviations; a behavior that confirms that sufficient time was provided before the sampling procedure for the simulated system to efficiently equilibrate. In addition, the magnitude of the fluctuations is extremely small, which confirms that the properties are accurately determined. It is worth noting that compressibility samples cannot be collected directly in the course of the simulation, in contrast to energy and pressure samples. In particular, the hypervirial theorem for $\mu_{\mathrm{ex}}$ involves the virial fluctuations, which can be evaluated only after the simulation is completed and the average virial is known. Therefore, in order to generate a set of compressibility samples which could be used for uncertainty analysis, we adopted the bootstrap resampling technique\,\cite{gould2006} that allowed us to construct $2048$ compressibility samples from the simulation data at each state point.

The resulting $u_{\mathrm{ex}}$, $p_{\mathrm{ex}}$ and $\mu_{\mathrm{ex}}$ values for all the roughly $150$ state points of interest have been tabulated in the supplementary material\,\cite{supplementary}. Both the average value and the standard deviation of each quantity are provided for the 2D-YOCP state points depicted in Fig.\ref{fig:phase_diagram}. All three thermodynamic properties obtained from the MD simulations are determined with a negligible statistical uncertainty that is quantified by a relative standard deviation (defined as the ratio between the standard deviation and the average value) which never exceeds $10^{-4}$.

It should be pointed out that, when the radial distribution function is known, then the reduced excess internal energy and reduced excess pressure can be computed from the following integral relations\,\cite{hansen2006}
\begin{align}
u_{\mathrm{ex}}(\Gamma,\kappa)&=\pi n\beta\int_0^\infty r\phi(r;\Gamma,\kappa)g(r;\Gamma,\kappa)dr\,,\label{eq:uex_rdf}\\
p_{\mathrm{ex}}(\Gamma,\kappa)&=-\frac{\pi n\beta}{2}\int_0^\infty r^2\frac{d\phi(r;\Gamma,\kappa)}{dr}g(r;\Gamma,\kappa)dr\,.\label{eq:pex_rdf}
\end{align}
This indirect extraction procedure was followed in earlier MD simulation works focusing on the thermodynamics of 2D-YOCP liquids\,\cite{hartmann2005,kryuchkov2017}. Our direct extraction procedure was preferred because, apart from being much faster, it allows for the quantification of statistical uncertainties in the determination of thermodynamic properties and does not suffer from tail or truncation errors.

\section{Equation of state}\label{sec:eos}

\noindent Practical equations of state specify the analytical relation between the reduced excess internal energy and the 2D-YOCP state variables, \emph{i.e.} $u_{\mathrm{ex}}(\Gamma,\kappa)$. Once the equation of state is determined, all other thermodynamic properties of the system follow from standard thermodynamic identities. In particular, for 2D-YOCP systems, an analytical expression for $u_{\mathrm{ex}}(\Gamma,\kappa)$ allows the computation of the reduced excess Helmholtz free energy from
\begin{equation}
f_{\mathrm{ex}}(\Gamma,\kappa)=\int_{0}^{\Gamma}\frac{u_{\mathrm{ex}}(\Gamma',\kappa)}{\Gamma'}d\Gamma'\,,\label{eq:fex_eos}
\end{equation}
the reduced excess pressure from
\begin{equation}
p_{\mathrm{ex}}(\Gamma,\kappa)=\frac{\Gamma}{2}\frac{\partial f_{\mathrm{ex}}(\Gamma,\kappa)}{\partial \Gamma}-\frac{\kappa}{2}\frac{\partial f_{\mathrm{ex}}(\Gamma,\kappa)}{\partial \kappa}\,,\label{eq:pex_eos}
\end{equation}
and the reduced excess inverse isothermal compressibility from
\begin{equation}
\mu_{\mathrm{ex}}(\Gamma,\kappa)=p_{\mathrm{ex}}(\Gamma,\kappa)+\frac{\Gamma}{2}\frac{\partial p_{\mathrm{ex}}(\Gamma,\kappa)}{\partial \Gamma}-\frac{\kappa}{2}\frac{\partial p_{\mathrm{ex}}(\Gamma,\kappa)}{\partial \kappa}\,.\label{eq:muex_eos}
\end{equation}

\subsection{Equations of state available in the literature}\label{sec:eos_old_eos}

\noindent In strongly coupled liquids, the reduced excess internal energy can be conveniently decomposed into the sum of two contributions\,\cite{hansen1973}: a static part $u_{\mathrm{st}}$ describing the energy of the system with its constituents frozen in a regular structure (at zero temperature) and a thermal part $u_{\mathrm{th}}$ accounting for the finite temperature effects that cause the particles to be displaced from such regular structure. The reduced excess internal energy decomposition reads as $u_{\mathrm{ex}}(\Gamma,\kappa)=u_{\mathrm{st}}(\Gamma,\kappa)+u_{\mathrm{th}}(\Gamma,\kappa)$ for the YOCP with the static component given by $u_{\mathrm{st}}(\Gamma,\kappa)=M(\kappa)\Gamma$ where $M(\kappa)$ is the Madelung constant.

For the 3D-YOCP, the Madelung constant is given by a simple closed-form expression\,\cite{rosenfeld1998} that can be obtained from the unitary packing fraction limit (also known as asymptotically high density limit) of the Percus-Yevick approximation for hard spheres\,\cite{wertheim1963,tolias2014} or the ion-sphere model\,\cite{khrapak2014,khrapak2015}. In addition, Rosenfeld and Tarazona (RT) have shown that the thermal component obeys the particularly simple scaling $u_{\mathrm{th}}(\Gamma,\kappa)\propto[\Gamma/\Gamma_\mathrm{m}(\kappa)]^{2/5}$ in the dense fluid region\,\cite{rosenfeld1998,rosenfeld2000} where the 3D-YOCP $\Gamma_\mathrm{m}(\kappa)$ is given by Eq.(4) of Ref.\cite{vaulina2002} and should not be confused with the 2D-YOCP $\Gamma_\mathrm{m}(\kappa)$ that is described by Eq.(\ref{eq:Gamma_melting}). A particularly attractive feature of the RT scaling has to do with its validity for a variety of three dimensional systems characterized by different interactions and molecular topology\,\cite{rosenfeld2000,lucco2019,ingebrigtsen2013a,ingebrigtsen2013b}. Furthermore, there exists a deep connection between isomorph theory and the RT scaling, with systems that follow the RT scaling often also being R-simple\,\cite{ingebrigtsen2013a}. In fact, it has been demonstrated that 3D-YOCP liquids are R-simple in an extensive region of their phase diagram\,\cite{veldhorst2015}. R-simple systems possess isomorph curves, \emph{i.e.} lines of constant excess entropy along which a large set of thermodynamic, structural and dynamic properties are approximately invariant when expressed in properly reduced units\,\cite{gnan2009,dyre2016}. The static part of the excess internal energy naturally produces no entropy, thus the excess entropy is exclusively computed from the thermal part of the excess internal energy via $s_{\mathrm{ex}}(\Gamma,\kappa)=u_{\mathrm{th}}(\Gamma,\kappa)-\int_{0}^{\Gamma}[u_{\mathrm{th}}(\Gamma',\kappa)/\Gamma']d\Gamma'$. Hence, the RT scaling is compatible with isomorph theory only if $\Gamma/\Gamma_{\mathrm{m}}(\kappa)$ is an accurate representation for the isomorphs. The latter is true for the 3D-YOCP\,\cite{veldhorst2015}, but in general the melting line constitutes an isomorphic line only to a first order approximation\,\cite{pedersen2016}.

For the 2D-YOCP, the situation is more complicated. The Madelung constant does not possess an analytical expression that can be derived from purely theoretical considerations, while the existence of a RT scaling for the thermal component is still open for debate and the functional form of the scaling is unknown. Moreover, no analytical representation is available for the 2D-YOCP isomorphs, neither is it even known whether the 2D-YOCP is R-simple. This lack of rigorous theoretical foundation for the construction of a 2D-YOCP equation of state has led, over the years, to the emergence of various functional forms for the analytical parametrization of the reduced excess internal energy.

Earlier attempts to obtain an analytical $u_{\mathrm{ex}}(\Gamma,\kappa)$ expression include the equation of state proposed by Hartmann and collaborators\,\cite{hartmann2005}, $u_{\mathrm{ex}}^{\mathrm{H}}(\Gamma,\kappa)=[a_\mathrm{H}(\kappa)+1/\kappa]\Gamma+b_\mathrm{H}(\kappa)\Gamma(\kappa)^{1/3}$ with the coefficients $a_{\mathrm{H}}(\kappa)$, $b_{\mathrm{H}}(\kappa)$ specified in Eqs.(3,4,7) of Ref.\cite{hartmann2005} as well as the equation of state presented by Vaulina\,\cite{vaulina2009} $u_{\mathrm{ex}}^{\mathrm{V}}(\Gamma,\kappa)=[a_\mathrm{V}(\kappa)+1/\kappa]\Gamma+b_\mathrm{V}(\kappa)$ with the coefficients $a_{\mathrm{V}}(\kappa)$, $b_{\mathrm{V}}(\kappa)$ specified below Eq.(4b) of Ref.\cite{vaulina2009}. In spite of a satisfactory accuracy for $\kappa\leq1.5$, these $u_{\mathrm{ex}}(\Gamma,\kappa)$ expressions have two major problems: they become very inaccurate for larger screening parameter values and do not lead to accurate thermodynamic properties via Eqs.(\ref{eq:pex_eos},\ref{eq:muex_eos}). Thus, in what follows, we focus on two more accurate $u_{\mathrm{ex}}(\Gamma,\kappa)$ expressions.

\emph{Kryuchkov and collaborators} have proposed the following $u_{\mathrm{ex}}(\Gamma,\kappa)$ equation of state that reads as\,\cite{kryuchkov2017}
\begin{equation}
	u_{\mathrm{ex}}^{\mathrm{K}}(\Gamma,\kappa)=M(\kappa)\Gamma+a_{\mathrm{K}}(\kappa)\ln\left[1+b_{\mathrm{K}}(\kappa)\Gamma^{s_{\mathrm{K}}(\kappa)}\right]\,,\label{eq:eos_kryuchkov}
\end{equation}
with $M(\kappa)$ the Madelung constant for 2D-YOCP crystals with triangular lattice that can be fitted with\,\cite{totsuji2004,kryuchkov2017}
\begin{align}
M(\kappa)=& -1.1061 +0.5038\kappa -0.11053\kappa^2 \nonumber\\
& + 0.00968\kappa^3 + 1/\kappa,\label{eq:eos_madelung}
\end{align}
while the unknown coefficients of the thermal component are $a_{\mathrm{K}}(\kappa)=0.357+0.094\kappa$, $b_{\mathrm{K}}(\kappa)=1.655\exp(-0.769\kappa)$, $s_{\mathrm{K}}(\kappa) = 0.688 - 0.052\kappa$. It is worth noting that an alternative fit for the thermal component was provided,\,where all the $\kappa$-dependence was absorbed in the form $\Gamma/\Gamma_\mathrm{m}(\kappa)$ with $\Gamma_{\mathrm{m}}(\kappa)$ given by Eq.(\ref{eq:Gamma_melting})\,\cite{kryuchkov2017}. The accuracy of the latter fit, which predicts $u_{\mathrm{ex}}$ within a few percent over the entire dense fluid region of the 2D-YOCP for $\kappa\leq3.0$\,\cite{kryuchkov2017}, supports the possibility of a modified RT scaling that is applicable to the 2D-YOCP.

Finally, \emph{Feng and co-workers} have proposed the following $u_{\mathrm{ex}}(\Gamma,\kappa)$ equation of state that reads as\,\cite{huang2017}
\begin{equation}
u_{\mathrm{ex}}^{\mathrm{F}}(\Gamma,\kappa)=a_{\mathrm{F}}(\kappa)\Gamma+b_{\mathrm{F}}(\kappa)\Gamma^{0.407},\,\label{eq:eos_feng}
\end{equation}
where the unknown coefficients for the static and thermal part are described by $a_{\mathrm{F}}(\kappa)=2(0.8394+0.5162\kappa)^{-6.4}$ and by $b_{\mathrm{F}}(\kappa)=2\exp(-1.579-0.3935\kappa)$.

Particular care should be taken during the application of these equations of state in the OCP limit $(\kappa=0)$. In this limit, it is necessary to replace the reduced excess internal energy $u_{\mathrm{ex}}$ with $u_{\mathrm{ex}}-\Gamma/\kappa$ in order to explicitly take into account the diverging background contribution, $\Gamma/\kappa$. It is evident that the equations of state proposed by Kryuchkov, Hartmann or Vaulina can be safely applied in the OCP limit by simply removing the $\Gamma/\kappa$ term, while the equation of state proposed by Feng and collaborators should not be applied in the OCP limit since the divergence is not removable.

\subsection{A new equation of state}\label{sec:eos_new_eos}

The equations of state for $u_{\mathrm{ex}}(\Gamma,\kappa)$ discussed in Section \ref{sec:eos_old_eos} were all obtained by fitting simulation data for the reduced excess internal energy alone. In what follows, a novel approach is presented that determines the equation of state for $u_{\mathrm{ex}}(\Gamma,\kappa)$ by simultaneously fitting simulation data for the reduced excess internal energy, pressure and inverse isothermal compressibility with the aid of the thermodynamic Eqs.(\ref{eq:fex_eos},\ref{eq:pex_eos},\ref{eq:muex_eos}). Initially, the internal energy is expressed as  $u_{\mathrm{ex}}(\Gamma,\kappa)=M(\kappa)\Gamma+u_{\mathrm{th}}(\Gamma,\kappa)$ with $M(\kappa)$ as given by Eq.\eqref{eq:eos_madelung} and the thermal component defined as
\begin{equation}
u_{\mathrm{th}}(\Gamma,\kappa)=a(\kappa)\Gamma+b(\kappa)\Gamma^{2/5}+c(\kappa)\Gamma\ln(\Gamma)\,.\label{eq:eos_my_uth}
\end{equation}
This parametrization is applicable in the OCP limit by simply removing the $\Gamma/\kappa$ term in the $M(\kappa)$ expression.

The first two terms in Eq.\eqref{eq:eos_my_uth} were inspired from the successful equation of state proposed by Hamaguchi and collaborators for 3D-YOCP liquids\,\cite{hamaguchi1996,hamaguchi1997} which contains terms proportional to $\Gamma$, $\Gamma^{s}$ and $\Gamma^{-s}$ with $s=1/3$. After trial and error, a different $s$ exponent was adopted and the $\Gamma^{-s}$ term had to be dropped, since it led to a non-monotonic pre-factor with respect to $\kappa$. The linear term acts as correction to the static component $M(\kappa)\Gamma$. The logarithmic term acts as a residual introduced to adjust the values for small coupling parameters $(\Gamma/\Gamma_{\mathrm{m}}(\kappa)\approx0.1)$ and was inspired from known low coupling expansions of the 3D-OCP internal energy in terms of $\Gamma\ln(\Gamma)$\,\cite{caillol2010}. The $\kappa$-dependent coefficients were expressed as Pade' approximants
\begin{align}
a(\kappa)&=\frac{a_{0}^{\mathrm{n}}+a_{1}^{\mathrm{n}}\kappa^{1/2}+a_{2}^{\mathrm{n}}\kappa+a_{3}^{\mathrm{n}}\kappa^2+a_{4}^{\mathrm{n}}\kappa^{5/2}}{1+a_{1}^{\mathrm{d}}\kappa^{1/2}+a_{2}^{\mathrm{d}}\kappa+a_{3}^{\mathrm{d}}\kappa^2+a_{4}^{\mathrm{d}}\kappa^{5/2}}\,,\label{eq:eos_my_aa}\\
b(\kappa)&=\frac{b_{0}^{\mathrm{n}}+b_{1}^{\mathrm{n}}\kappa^{1/2}+b_{2}^{\mathrm{n}}\kappa+b_{3}^{\mathrm{n}}\kappa^2+b_{4}^{\mathrm{n}}\kappa^{5/2}}{1+b_{1}^{\mathrm{d}}\kappa^{1/2}+b_{2}^{\mathrm{d}}\kappa+b_{3}^{\mathrm{d}}\kappa^2+b_{4}^{\mathrm{d}}\kappa^{5/2}}\,,\label{eq:eos_my_bb}\\
c(\kappa)&=\frac{c_{0}^{\mathrm{n}}+c_{1}^{\mathrm{n}}\kappa^{1/2}+c_{2}^{\mathrm{n}}\kappa}{1+c_{1}^{\mathrm{d}}\kappa^{1/2}+c_{2}^{\mathrm{d}}\kappa}\,.\label{eq:eos_my_cc}
\end{align}
Alternative expressions for the Pade' approximants containing only powers of $\kappa$ or of $\kappa^{1/4}$ were also tested, but proved to be less accurate than the above approximants.

\begin{table}[t]
	\caption{Numerical coefficients for the Pade' approximants of Eqs.(\ref{eq:eos_my_aa},\ref{eq:eos_my_bb},\ref{eq:eos_my_cc}) that determine the $\kappa$-dependent coefficients $a(\kappa)$, $b(\kappa)$, $c(\kappa)$ appearing in the parametrization of the thermal part of the reduced excess internal energy, see Eq.\eqref{eq:eos_my_uth}.}
	\label{tab:eos_coefficients}
	\renewcommand\arraystretch{1.5}
	\centering
	\begin{tabular}{cccccc}
		\toprule
		 & $i=0$ & $i=1$ & $i=2$ & $i=3$ & $i=4$ \\
		\hline
		$a_{i}^{\mathrm{n}}$ & -0.022587 & -40.935& 43.6611& -12.2860& 3.22385 \\
		$a_{i}^{\mathrm{d}}$ & --- & 1569.37& -1524.38&
		553.496& -188.020 \\
		$b_{i}^{\mathrm{n}}$ & 0.361510 & 3.53190& -3.43696& 0.864783& -0.224233 \\
		$b_{i}^{\mathrm{d}}$ & --- & 9.46041& -8.99282&
		3.02666& -0.897450 \\
		$c_{i}^{\mathrm{n}}$ & 0.002812 & -0.003326 & 0.000993 & --- & --- \\
		$c_{i}^{\mathrm{d}}$ & --- & -1.17692 & 0.442468 & --- & --- \\
		\hline
	\end{tabular}
	\renewcommand\arraystretch{1.0}
\end{table}

The OCP coefficients $a_0^{\mathrm{n}}$, $b_0^{\mathrm{n}}$ and $c_0^{\mathrm{n}}$ were determined by fitting the OCP simulation results for the thermal component of the reduced excess internal energy that have been tabulated in Table II of Ref.\cite{gann1979}. The remaining coefficients were determined as follows. Starting from the reduced excess internal energy from MD simulations, $u_{\mathrm{ex}}^{\mathrm{MD}}$, the thermal component of the excess internal energy was computed as $u_{\mathrm{th}}^{\mathrm{MD}}=u_{\mathrm{ex}}^{\mathrm{MD}}-M(\kappa)\Gamma$. Then, $u_{\mathrm{th}}^{\mathrm{MD}}$ was fitted with Eq.(\ref{eq:eos_my_uth}) six times, one for each value of $\kappa_i$ belonging to the set $\kappa=\{0.5,1.0,1.5,2.0,2.5,3.0\}$ and for all corresponding $\Gamma$ values depicted in Fig.\ref{fig:phase_diagram}. The resulting six values of $c(\kappa_i)$ were fitted with the Pade' approximant given in Eq.(\ref{eq:eos_my_cc}) to define the coefficients $c_{1}^{\mathrm{n}},c_{2}^{\mathrm{n}},c_{3}^{\mathrm{n}}, c_{1}^{\mathrm{d}}$ and $c_{2}^{\mathrm{d}}$. The resulting six values for $a(\kappa_i)$, $b(\kappa_i)$ were stored for later analysis. Afterwards, the leading contribution to the thermal pressure $ p_{\mathrm{th,l}}^{\mathrm{MD}}=p_{\mathrm{ex}}^{\mathrm{MD}}-p_{\mathrm{st}}(\Gamma,\kappa)-\delta p(\Gamma,\kappa)$ and the leading contribution to the thermal inverse compressibility $\mu_{\mathrm{th,l}}^{\mathrm{MD}}=\mu_{\mathrm{ex}}^{\mathrm{MD}}-\mu_{\mathrm{st}}(\Gamma,\kappa)-\delta \mu(\Gamma,\kappa)$ were computed. The thermodynamic Eqs.(\ref{eq:fex_eos},\ref{eq:pex_eos},\ref{eq:muex_eos}) were employed to obtain the static components $p_{\mathrm{st}}(\Gamma,\kappa)$ and $\mu_{\mathrm{st}}(\Gamma,\kappa)$ from $u_{\mathrm{st}}(\Gamma,\kappa)=M(\kappa)\Gamma$ and the residual components $\delta p(\Gamma,\kappa)$ and $\delta\mu(\Gamma,\kappa)$ from $\delta u_{\mathrm{st}}(\Gamma,\kappa)=c(\kappa)\Gamma\ln(\Gamma)$. On the other hand, by applying the thermodynamic Eqs.(\ref{eq:fex_eos},\ref{eq:pex_eos},\ref{eq:muex_eos}) to the leading component of the fit for the excess internal energy, $a(\kappa)\Gamma+b(\kappa)\Gamma^{2/5}$, we obtained that the leading contribution to the thermal pressure could be expressed as
\begin{equation}
	p_{\mathrm{th,l}}(\Gamma,\kappa)=a_p(\kappa)\Gamma+b_p(\kappa)\Gamma^{2/5}\label{eq:eos_my_pth}
\end{equation}
and that the leading contribution to the thermal inverse isothermal compressibility could be parameterized with
\begin{equation}
	\mu_{\mathrm{th,l}}(\Gamma,\kappa)=a_\mu(\kappa)\Gamma+b_\mu(\kappa)\Gamma^{2/5}.\,\label{eq:eos_my_muth}
\end{equation}
The four $\kappa$-dependent coefficients in Eqs.(\ref{eq:eos_my_pth},\ref{eq:eos_my_muth}) are connected to the $a(\kappa)$, $b(\kappa)$ coefficients in Eq.(\ref{eq:eos_my_uth}) via
\begin{align}
a_p(\kappa)&=\frac{1}{2}\left[a(\kappa)-\kappa\frac{da(\kappa)}{d\kappa}\right]\,,\label{eq:eos_my_ap}\\
b_p(\kappa)&=\frac{1}{2}\left[b(\kappa)-\frac{5}{2}\kappa\frac{db(\kappa)}{d\kappa}\right]\,,\label{eq:eos_my_bp}\\
a_\mu(\kappa)&=\frac{3}{4}\left[a(\kappa)-\kappa\frac{da(\kappa)}{d\kappa}+\frac{1}{3}\kappa^2\frac{d^2a(\kappa)}{d\kappa^2}\right]\,,\label{eq:eos_my_am}\\
b_\mu(\kappa)&=\frac{3}{5}\left[b(\kappa)-\frac{15}{8}\kappa\frac{db(\kappa)}{d\kappa}+\frac{25}{24}\kappa^2\frac{d^2b(\kappa)}{d\kappa^2}\right]\,.\label{eq:eos_my_bm}
\end{align}
Therefore, $p_{\mathrm{th,l}}^{\mathrm{MD}}$ and $\mu_{\mathrm{th,l}}^{\mathrm{MD}}$ were fitted with Eqs.(\ref{eq:eos_my_pth},\ref{eq:eos_my_muth}) for six $\kappa_i$ belonging to $\kappa=\{0.5,1.0,1.5,2.0,2.5,3.0\}$ producing twenty-four values for $a_p(\kappa_i)$, $b_p(\kappa_i)$, $a_\mu(\kappa_i)$ and $b_\mu(\kappa_i)$, six for each type of coefficient. Finally, the sets of coefficients $\{a_{1}^{\mathrm{n}},a_{2}^{\mathrm{n}},a_{3}^{\mathrm{n}},a_{4}^{\mathrm{n}}\}$ and $\{a_{1}^{\mathrm{d}},a_{2}^{\mathrm{d}},a_{3}^{\mathrm{d}},a_{4}^{\mathrm{d}}\}$, were determined by simultaneously fitting  the six values for $a(\kappa_i)$ with Eq.\eqref{eq:eos_my_aa}, the six values for $a_p(\kappa_i)$ with Eq.\eqref{eq:eos_my_ap} and the six values for $a_\mu(\kappa_i)$ with Eq.\eqref{eq:eos_my_am}. An analogous procedure was adopted to define the sets of coefficients  $\{b_{1}^{\mathrm{n}},b_{2}^{\mathrm{n}},b_{3}^{\mathrm{n}},b_{4}^{\mathrm{n}}\}$ and $\{b_{1}^{\mathrm{d}},b_{2}^{\mathrm{d}},b_{3}^{\mathrm{d}},b_{4}^{\mathrm{d}}\}$ which were found by simultaneously fitting the eighteen values for $b(\kappa_i)$,$b_p(\kappa_i)$ and $b_\mu(\kappa_i)$ with Eqs.(\ref{eq:eos_my_bb},\ref{eq:eos_my_bp},\ref{eq:eos_my_bm}). The coefficients are summarized in Table \ref{tab:eos_coefficients}.

\begin{table*}[htbp]
	\caption{Mean absolute relative deviations between the predictions of the new equation of state Eq.(\ref{eq:eos_my_uth}) (superscript N), the Kryuchkov and collaborators equation of state Eq.(\ref{eq:eos_kryuchkov}) (superscript K), the Feng and co-workers equation of state Eq.(\ref{eq:eos_feng}) (superscript F) and the results of the MD simulations discussed in Section \ref{sec:md_thermo}. The average thermodynamic quantity deviation for each screening parameter was computer over all the corresponding coupling parameters, see the summary of  Fig.\ref{fig:phase_diagram}. Deviations for the reduced excess internal energy are reported in columns 2-4 ($\epsilon_{\mathrm{u}}$), deviations for the reduced excess pressure are reported in columns 5-7 ($\epsilon_{\mathrm{p}}$) and deviations for the reduced excess inverse isothermal compressibility are provided in columns 8-10 ($\epsilon_{\mathrm{\mu}}$).}\label{tab:eos_errors}
	\renewcommand\arraystretch{1.5}
	\centering
	\begin{tabular}{cccccccccc}
		\toprule
		$\kappa$ & $\epsilon^{\mathrm{N}}_{\mathrm{u}} (\%)$ & $\epsilon^{\mathrm{K}}_{\mathrm{u}}(\%)$ & $\epsilon^{\mathrm{F}}_{\mathrm{u}}$(\%) & $\epsilon^{\mathrm{N}}_{\mathrm{p}}$(\%) & $\epsilon^{\mathrm{K}}_{\mathrm{p}}$(\%) & $\epsilon^{\mathrm{F}}_{\mathrm{p}}$(\%) & $\epsilon^{\mathrm{N}}_{\mathrm{\mu}}$(\%) & $\epsilon^{\mathrm{K}}_{\mathrm{\mu}}$(\%) & $\epsilon^{\mathrm{F}}_{\mathrm{\mu}}$(\%)\\
		\hline
		0.5	&	0.035	&	0.049	&	0.531	&	0.013	&	0.122	&	4.563	&	0.009	&	0.087	&	11.545	\\
		1.0	&	0.130	&	0.051	&	1.850	&	0.132	&	0.124	&	1.205	&	0.019	&	0.185	&	3.485	\\
		1.5	&	0.276	&	0.067	&	3.018	&	0.288	&	0.279	&	4.866	&	0.054	&	0.094	&	2.673	\\
		2.0	&	0.197	&	0.531	&	3.949	&	0.321	&	0.063	&	5.929	&	0.242	&	0.846	&	7.706	\\
		2.5	&	1.008	&	0.365	&	19.619	&	0.524	&	2.337	&	0.170	&	0.322	&	0.589	&	7.272	\\
		3.0	&	1.026	&	0.707	&	44.510	&	0.489	&	4.658	&	12.484	&	0.240	&	19.472	&	0.717	\\
		\hline
	\end{tabular}
	\renewcommand\arraystretch{1.0}
\end{table*}

\subsection{Level of accuracy of different equations of state}\label{sec:eos_compare_eos}

\noindent The predictions of the new equation of state, see Eq.(\ref{eq:eos_my_uth}), and the two recent literature equations of state, see Eqs.(\ref{eq:eos_kryuchkov},\ref{eq:eos_feng}), have been extensively compared against the MD thermodynamic property results presented in Section \ref{sec:md_thermo} and tabulated in the supplementary material\,\cite{supplementary}. The summary of this comparison is reported in Table \ref{tab:eos_errors}.

Concerning the reduced excess internal energy, our new equation of state and the Kryuchkov equation of state are visibly more accurate than the Feng equation of state. In particular, the Feng equation leads to exceptionally large errors for  $\kappa>2.0$, while the other two equations are both able to predict $u_{\mathrm{ex}}$ within $1\%$ over the entire $\kappa\leq3.0$ range, with the Kryuchkov equation of state having a slight edge. Concerning the reduced excess pressure, the situation is somewhat similar with the important difference that the present equation of state is accurate within $0.5\%$ for any value of $\kappa$, whereas the performance of the Kryuchkov equation of state abruptly degrades when $\kappa\geq2.5$. Concerning the reduced excess inverse isothermal compressibility, the new equation of state remains superior being accurate within $0.3\%$ for any value of $\kappa$, while the Kryuchkov equation of state has a high accuracy up to $\kappa=2.5$, but it becomes accurate only within $\sim20\%$ at $\kappa=3.0$.

Overall, it is concluded that the new equation of state proposed in Section \ref{sec:eos_new_eos} leads to improvements over all the other dense 2D-YOCP liquid equations of state currently available in the literature, especially when it comes to predictions of thermodynamic properties for $\kappa>2.0$. Nevertheless, it must be noted that the equation of state proposed by Kryuchkov and collaborators exhibits an excellent agreement with MD simulations, despite possessing a simple $\kappa$-dependence for the thermal component. In light of such good agreement, it would seem reasonable to replace Eq.\eqref{eq:eos_my_uth} with $u_{\mathrm{th}}(\Gamma,\kappa)=\tilde{a}(\kappa)\ln[1+\tilde{b}(\kappa)^{\tilde{s}(\kappa)}]$. This possibility was tested but eventually discarded because the coefficients $\tilde{a}(\kappa),\tilde{b}(\kappa)$ and $\tilde{s}(\kappa)$ showed un unfavorable dependence over $\kappa$ characterized by changes of sign and stationary points.

\section{Integral equation theory}\label{sec:iet}

\subsection{Method}\label{sec:iet_theory}

\noindent For an isotropic pair-interacting one-component system, the integral equation theory of liquids enables the computation of two-particle equilibrium correlation functions by combining the Ornstein-Zernike integral equation\,\cite{hansen2006}
\begin{equation}
h(r)=c(r)+n\int c(r')h(|\boldsymbol{r}-\boldsymbol{r}'|)d^2r'\,\label{eq:theory_oz}
\end{equation}
with the formally exact non-linear closure equation\,\cite{hansen2006}
\begin{equation}
g(r)=\exp\left[-\beta u(r)+h(r)-c(r)+B(r)\right]\,.\label{eq:theory_oz_closure}
\end{equation}
In the above, $h(r)=g(r)-1$ is the total correlation function and $c(r)$ is the direct correlation function. An expression for the bridge function $B(r)$ is necessary to complete the theory. It is generally prescribed by approximations that are constructed on theoretical grounds\,\cite{hansen2006,bomont2008} or that are aided by computer simulations\,\cite{iyetomi1992,tolias2019}. Popular approximations include the hypernetted-chain (HNC) approach which assumes that $B(r)=0$ and has proven to be successful for systems with soft interaction potentials\,\cite{ng1974,murillo2003,yazdi2014,yazdi2015} or the Percus Yevick approach which assumes that $B(r)=\ln[1+\gamma(r)]-\gamma(r)$ and has proven to be successful for hard-sphere-like systems\,\cite{lado1968,chae1969}. Here $\gamma(r)=h(r)-c(r)$ is the indirect correlation function.

When combined with a $B(r)$ assumption, Eqs.(\ref{eq:theory_oz},\ref{eq:theory_oz_closure}) form a system of equations to be solved for $g(r)$. For its numerical solution, a validated algorithm was followed that was previously applied to three-dimensional systems \cite{tolias2019,lucco2021c}. It is based on Picard iterations in Fourier space combined with a standard mixing technique\,\cite{ng1974} and a long-range decomposition method\,\cite{lado1978} in the OCP limit. The convergence criterion was formulated in terms of the Fourier transform of the indirect correlation function, $\gamma(k)$, and chosen to be $||\gamma_m(k)-\gamma_{m-1}(k)||<10^{-5}\ \forall k$. The two-dimensional Fourier transforms were first expressed as one-dimensional Hankel transforms and were then computed with the Quasi-Fast Hankel Transform algorithm\,\cite{siegman1977} over a discrete grid of $N_{\mathrm{p}}$ points. The chosen algorithm allowed to circumvent the unfavorable $O(N_{\mathrm{p}}^2)$ scaling of direct Hankel transform calculations and did not feature the long-wavelength deficiencies which were observed for similar algorithms that were adopted in earlier works\,\cite{talman1978,caillol1981,hansen1981}. Nevertheless, it required the introduction of a computational grid with a constant logarithmic spacing both in real and Fourier space. We employed a grid of $N_{\mathrm{p}}=32768$ points that extended from $3.5\times10^{-6}d$ up to $50d$ in real space and from $3.5\times10^{-6}/d$ up to $50/d$ in Fourier space, with both grids featuring the same logarithmic spacing, i.e. $\log(r_i/r_{i-1} )=\log(k_i/k_{i-1})=5\times10^{-4}$. The algorithm was successfully benchmarked against HNC results for the 2D-YOCP that are available in the literature\,\cite{lado1978,murillo2003}.

\subsection{Scaled HNC approach}\label{sec:iet_shnc}

\begin{figure}[t]
	\centering
	\includegraphics[width=3.4in]{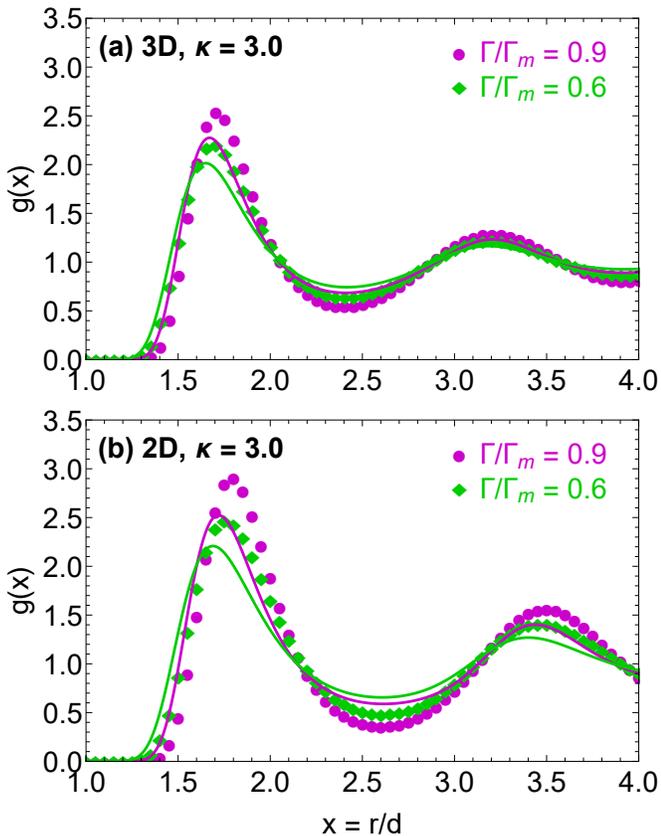}
	\caption{Radial distribution functions for the 3D-YOCP (panel a) and the 2D-YOCP (panel b) as obtained from MD simulations (discrete points) and the HNC approach (solid lines). Results for two state points characterized by $\kappa=3.0$ and two values of the normalized coupling parameter $\Gamma/\Gamma_{\mathrm{m}}$, namely $\Gamma/\Gamma_{\mathrm{m}}=0.6$ (green) and $\Gamma/\Gamma_{\mathrm{m}}=0.9$ (magenta). For the 3D-YOCP,  $\Gamma_\mathrm{m}$ is given in Eq.(4) of Ref.\cite{vaulina2002} and $d=(4\pi n/3)^{-1/3}$. For the 2D-YOCP, $\Gamma_{\mathrm{m}}$ is given by Eq.(\ref{eq:Gamma_melting}) and $d=(\pi n)^{-1/2}$.}\label{fig:rdf_MD_HNC_2D_3D}
\end{figure}

\noindent The HNC approach possesses a reasonable accuracy for the 3D-YOCP, being able to reproduce the main features of the radial distribution function with an accuracy of $\sim20\%$ and the thermodynamic properties within $5\%$\,\cite{tolias2019}. However, the 2D-YOCP is known to be richer in structure than its three-dimensional counterpart, which translates to a decline in the accuracy of the HNC predictions. This is demonstrated in Fig.\ref{fig:rdf_MD_HNC_2D_3D}, where the MD-extracted and HNC-generated radial distribution functions are illustrated for constant $(\kappa,\,\Gamma/\Gamma_\mathrm{m})$ pairs in the 3D and 2D case. It is evident that the maxima and minima of the radial distribution function become more pronounced and that the deviations between MD results and HNC predictions become larger as the dimensionality decreases.

In spite of these deficiencies of the HNC approach, no advanced integral equation theory approximations have been developed that would lead to more accurate structural predictions for 2D-YOCP liquids. This is in stark contrast to 3D-YOCP liquids, for which two very accurate approximations are available: namely the IEMHNC approach based on the isomorph invariance property of the bridge functions of R-simple systems\,\cite{tolias2019,lucco2021a} and the VMHNC approach based on the notion of bridge function quasi-universality\,\cite{rosenfeld1979,rosenfeld1986}. Such advanced approaches are characterized by an accuracy of $<2\%$ within the first coordination cell of the radial distribution function\,\cite{lucco2021c}. Unfortunately, the IEMHNC and VMHNC approach are not directly applicable to the 2D-YOCP: the IEMHNC approach because of the lack of a parameterized 2D-OCP bridge function to be used for the construction of the 2D-YOCP bridge function with the aid of the isomorph mapping\,\cite{tolias2019} and the VMHNC approach because of the lack of a reference system with a known analytical solution to be used for the construction of the VMHNC free energy functional\,\cite{rosenfeld1986}. It is worth to mention the crossover approach of Ref.\cite{ballone1985}, which constitutes a notable attempt to improve the HNC accuracy for the 2D-OCP.

In light of the above, we opted to take advantage of the vast corpus of MD results presented in Section \ref{sec:md_thermo} in order to construct an empirical correction to the HNC approach which improves its predictions without requiring any additional input other than that already available. The sought-for correction was constructed by considering that: (a) the main shortcoming of the HNC approach within the first coordination cell refers to the large underestimation of the magnitude of the global $g(r)$ maximum, see the lower panel of Fig.\ref{fig:rdf_MD_HNC_2D_3D}, (b) our earlier 3D-YOCP work has demonstrated that the HNC approach produces highly accurate structural properties provided that the state point is rescaled towards the stronger coupling region\,\cite{lucco2021c}. Given the above, our scaled hypernetted-chain approach (SHNC) was based on retaining the assumption of a vanishing bridge function provided that the interaction strength is up-scaled in a manner that reproduces the exact first peak of $g(r)$. In other words, a mapping is employed from the actual state point $(\Gamma,\kappa)$ to another state point $(\Gamma_{\mathrm{SHNC}}\geq\Gamma,\kappa)$ with the unknown $\Gamma_{\mathrm{SHNC}}$ determined by the condition that the HNC approach at $(\Gamma_{\mathrm{SHNC}},\kappa)$ leads to the MD-extracted first peak of $g(r)$ at $(\Gamma,\kappa)$. The idea of interaction strength rescaling within the HNC approach dates back to the seminal work of Ng for the 3D-OCP\,\cite{ng1974}, whereas an oversimplified version of this idea is encountered in the recent T/2-HNC approach proposed for supercooled dipolar binary mixtures\,\cite{hajnal2011}.

The SHNC approach is based on the implicit assumption that tempering with the HNC interaction strength in order to ensure an exact global maximum magnitude does not have a detrimental effect on other features of the radial distribution function, especially within the first coordination cell. Since there is no rigorous way of justifying such assumption, the discussion on its validity is postponed to Section \ref{sec:iet_rdf}, where the predictions of the SHNC approach are compared with "exact" MD results.

In integral equation theory, the bridge function and dimensionless interaction potential appear only in the closure equation combined as $\beta u(r)-B(r)$. Therefore, our tempering of the interaction potential within the HNC approach is equivalent to approximating the bridge function with $B(x;\Gamma,\kappa)=\left[\Gamma-\Gamma_{\mathrm{SHNC}}(\Gamma,\kappa)\right]\exp(-\kappa x)/x$ or with $B(x;\Gamma,\kappa)=\left\{\left[\Gamma-\Gamma_{\mathrm{SHNC}}(\Gamma,\kappa)\right]/\Gamma\right\}\beta{u(x)}$. The unknown function $\Gamma_{\mathrm{SHNC}}(\Gamma,\kappa)$ is obtained in the following manner: (a) for each $(\Gamma,\kappa)$ state point, the coupling parameter is gradually up-scaled and the HNC approach is numerically solved until the first $g(r)$ peak coincides with the respective MD result, (b) this procedure is repeated for all the $(\Gamma,\kappa)$ state points considered in the computer simulations reported in Section \ref{sec:md_parameters} as well as for all the OCP state points simulated in Ref.\cite{ott2015} and a dataset for $(\Gamma,\kappa,\Gamma_{\mathrm{SHNC}})$ is generated, (c) a closed-form expression for $\Gamma_{\mathrm{SHNC}}(\Gamma,\kappa)$ is acquired by sequential least-square fitting with respect to $\Gamma,\,\kappa$. The SHNC mapping reads as
\begin{equation}
\Gamma_{\mathrm{SHNC}}(\Gamma,\kappa)=\Gamma + p(\kappa)\Gamma^{3/2} + q(\kappa)\Gamma^{2}\,\label{eq:shnc_mapping}
\end{equation}
where
\begin{align}
p(\kappa)&=\frac{0.17013 - 0.03498\kappa + 0.00157\kappa^2}{1 - 0.25448\kappa + 0.36940\kappa^2}\,,\\
q(\kappa)&=\frac{-0.00572 + 0.003012\kappa - 0.00044\kappa^2}{1 - 0.63152\kappa + 0.57005\kappa^2}\,.
\end{align}

It is important to emphasize that, because the mapping of Eq.(\ref{eq:shnc_mapping}) was obtained by fitting, the SHNC approach should not be extrapolated beyond the original fitting region of $\Gamma/\Gamma_{\mathrm{m}}(\kappa)\in[0.1,1.0]$ and $\kappa\in[0.0,3.0]$. However, since the fit was constructed in such a way that the SHNC approach reduces to the HNC approach for $\Gamma\rightarrow0$, weak coupling extrapolations are permissible. Therefore, it can be concluded that the 2D-YOCP phase diagram region of validity of the SHNC approach is $\Gamma/\Gamma_{\mathrm{m}}(\kappa)\leq1.0$ and $\kappa\leq3.0$. In other words, provided that the screening parameter is not large, the SHNC can be employed in the entire stable fluid region but not for metastable states.

It is worth noting that we explored the possibility to construct the SHNC mapping by taking advantage of the effective coupling parameter introduced in Ref.\cite{hartmann2005}, $\Gamma^*(\Gamma,\kappa)=\Gamma f(\kappa)$, with $f(\kappa)=1-0.388\kappa^2 + 0.138\kappa^3 - 0.0138\kappa^4$. This effective coupling parameter is related to the melting line parametrization of Eq. \eqref{eq:Gamma_melting} which, in fact, can be also expressed via $\Gamma_{\mathrm{m}}(\kappa)=\Gamma^*(\Gamma_{\mathrm{m}}^{\mathrm{OCP}},\kappa)/f^2(\kappa)$. Since the magnitude of the first maximum of $g(r)$ is approximately constant for state points with the same $\Gamma^*(\Gamma,\kappa)$\,\cite{hartmann2005}, it should have been possible to construct an SHNC approach where $\Gamma^*(\Gamma,\kappa)$ is employed to map any YOCP state point to an effective OCP state point $(\Gamma^*,\kappa=0)$ which is then rescaled so that the HNC result for the first $g(r)$ peak coincides with the results of MD simulations. Such a mapping was tested, but was eventually discarded because it showed pronounced deviations from the simulation results in the region characterized by $\Gamma/\Gamma_{\mathrm{m}}(\kappa)\leq0.2$ and $\kappa=3.0$ that were traced back to inaccuracies in the parametrization of the effective coupling parameter $\Gamma^*(\Gamma,\kappa)$\,\cite{hartmann2005}.

\subsection{Structural properties}\label{sec:iet_rdf}

\noindent The HNC and SHNC approaches were numerically solved for all the 2D-YOCP state points illustrated in Fig.\ref{fig:phase_diagram} and all the 2D-OCP state points simulated in Ref.\cite{ott2015}. The computed radial distribution functions have been compared to the ones extracted from computer simulations.

\begin{figure}[t!]
	\centering
	\includegraphics[width=3.4in]{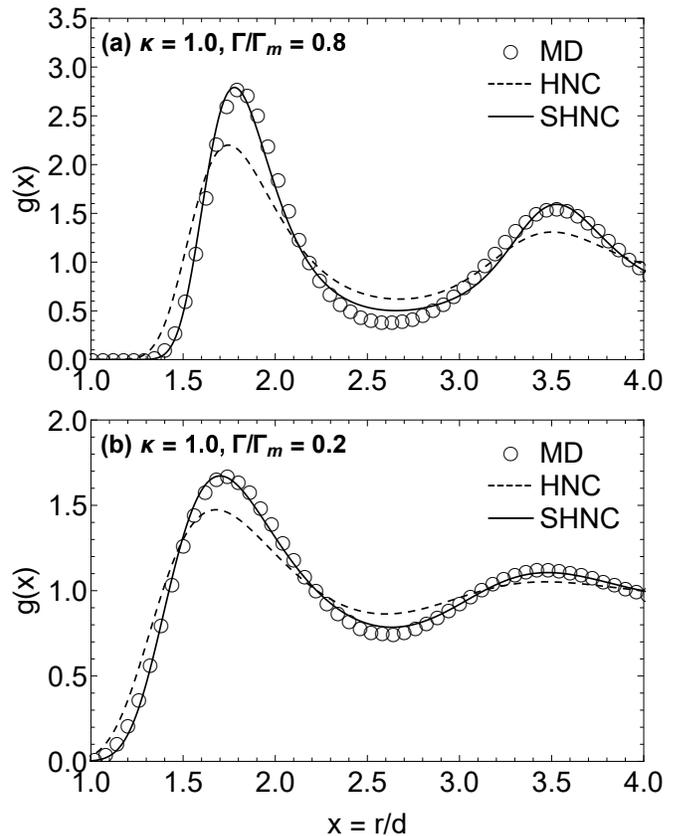}
	\caption{Radial distribution functions of 2D-YOCP liquids acquired from MD simulations (discrete points), the HNC approach (dashed lines) and the SHNC approach (solid lines). Results for two state points at the opposite ends of the dense fluid region, namely $(\kappa=1.0, \Gamma/\Gamma_{\mathrm{m}}=0.8)$ in panel (a) and $(\kappa=1.0, \Gamma/\Gamma_{\mathrm{m}}=0.2)$  in panel (b). Recall that the analytical approximation for the melting line, $\Gamma_{\mathrm{m}}(\kappa)$, is given by Eq.(\ref{eq:Gamma_melting}).}\label{fig:rdf_MD_HNC_SHNC}
\end{figure}

As illustrated in Fig.\ref{fig:rdf_MD_HNC_SHNC}, multiple advantages are gained by adopting the SHNC in place of the HNC approach. In fact, modification of the HNC interaction strength to ensure an exact first peak magnitude has a positive effect on all features of the radial distribution function within the first and second coordination cells. To be more specific, apart from the expected enormous improvement concerning the magnitude of the first peak, there is also a strong improvement in the correlation void, the magnitude of the first trough and the magnitude of the second peak as well as a slight improvement in the positions of all peaks and troughs.

\begin{figure}[t]
	\centering
	\includegraphics[width=3.4in]{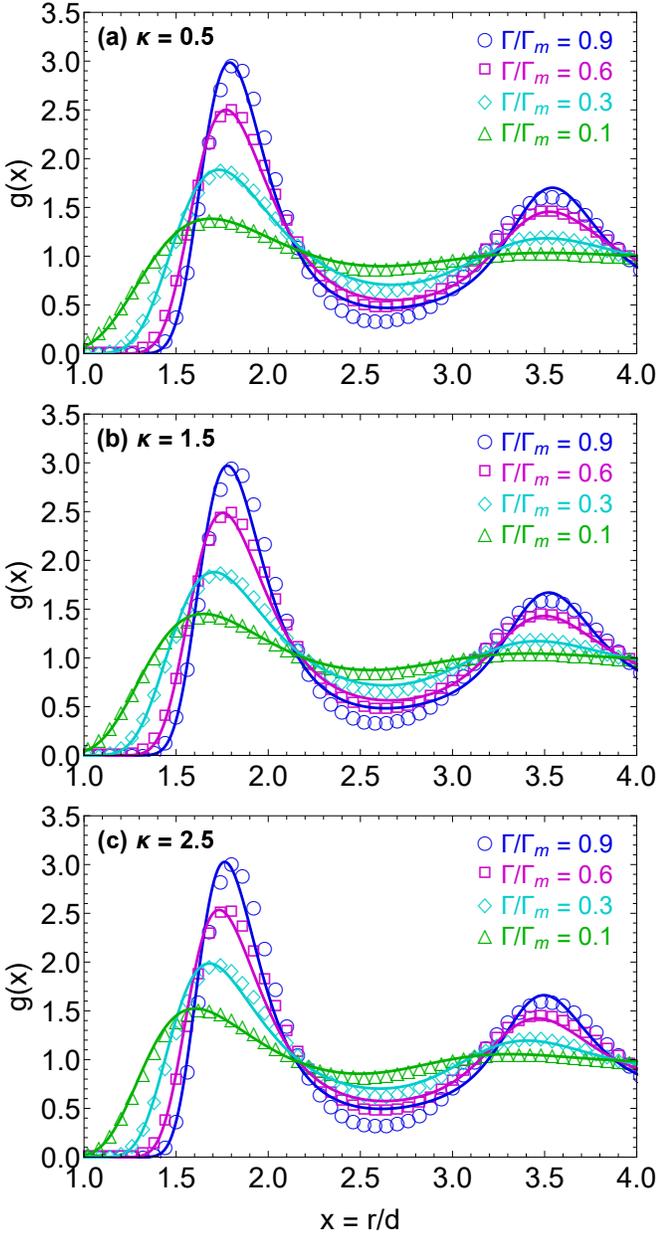}
	\caption{Radial distribution functions of 2D-YOCP liquids acquired from MD simulations (discrete points) and the SHNC approach (solid lines). Each panel focuses on a single screening parameter ($\kappa=0.5$ in a, $\kappa=1.5$ in b, $\kappa=2.5$ in c) and four coupling parameters $\Gamma/\Gamma_{\mathrm{m}}=0.1,\,0.3,\,0.6,\,0.9$ color-coded by green, cyan, magenta and blue, respectively.}\label{fig:rdf_MD_SHNC}
\end{figure}

The superior performance of the SHNC approach compared to the HNC approach is already evident at small coupling, but it becomes much more pronounced in the vicinity of the melting line. As demonstrated in Fig.\ref{fig:rdf_MD_SHNC}, the SHNC approach is capable of producing accurate predictions for the radial distribution function throughout the whole dense fluid region of the 3D-YOCP phase diagram.

For a more quantitative assessment of the SHNC and HNC accuracy, we report the relative deviations in key $g(r)$ figures of merit: the location of the edge of the correlation void (assumed to be given by the first location where $g(r/d)=0.5$) as well as the magnitudes and positions of the first maximum, first non-zero minimum and second maximum. The values of these quantities for "exact" radial distribution functions are available in Table 1 of Ref.\cite{ott2015} for $\kappa=0.0$ and in the supplementary material\,\cite{supplementary} for $0.5\leq\kappa\leq3.0$. The most noticeable SHNC improvements take place in the location of the correlation void and in the magnitude of the first maximum. The former is predicted within $1\%$ from the SHNC approach and only within 5\% from the HNC approach. The latter is predicted within $0.5\%$ from the SHNC approach, whereas it is consistently strongly underestimated in the HNC approach with the relative deviations even exceeding $25\%$ close to the melting line. Concerning the locations of the first and the second maximum as well as of the first non-zero minimum, the SHNC slightly improves the HNC predictions which are anyways well within $5\%$. Furthermore, both approaches produce rather poor estimates for the magnitude of the first non-zero minimum, but the HNC approach is much more accurate than the SHNC approach with $20\%$ vs $40\%$ mean relative deviations from the MD results. Finally, concerning the magnitude of the second maximum, the SHNC is accurate within $\sim2\%$ and the HNC only within $\sim10\%$.

\subsection{Thermodynamic properties}\label{sec:iet_therm}

\noindent The performance of the HNC and the SHNC integral equation theory approximations was also evaluated at the level of the thermodynamic properties. For this purpose, Eqs.(\ref{eq:uex_rdf},\ref{eq:pex_rdf},\ref{eq:muex_eos}) were employed to compute the excess internal energy, excess pressure and excess inverse isothermal compressibility from the radial distribution functions obtained with the two approaches. The results were then compared with the thermodynamic properties that were extracted from MD simulations and tabulated in the supplementary material\,\cite{supplementary}.

Near the OCP limit, both approaches are able to reproduce all the three thermodynamic properties within 1\%, namely the SHNC within $\sim0.2\%$ and the HNC within $\sim0.6\%$. However, the accuracy of the SHNC approach remains almost constant with the screening parameter, whereas the performance of HNC approach promptly degrades as the screening parameter increases. Concerning the excess internal energy, the mean deviations reach $2\%$ for the SHNC and $10\%$ for the HNC for $\kappa=3$. Concerning the excess pressure, the mean deviations reach $2\%$ for the SHNC and $8\%$ for the HNC for $\kappa=3$. Concerning the excess inverse isothermal compressibility, the mean deviations remain $<1\%$ for the SHNC and reach $5\%$ for the HNC for $\kappa=3$. To sum up, it can be concluded that the SHNC approach reproduces the excess internal energy, pressure and isothermal compressibility within $2\%$ over the entire dense fluid region, while the HNC approach produces estimates which are accurate within $10\%$.

It is important to point out that the virial route was followed for the computation of the excess inverse isothermal compressibility. The utilization of the so-called statistical route, \emph{i.e.} $\mu_{\mathrm{ex}}=-{n}\int{c}(r)d^2r$\,\cite{hansen2006}, would result to large deviations from the exact results for both approximations and especially for the SHNC approach. The SHNC bridge function does not obey the correct asymptotic limit $B(r)\to-{h}^2(r)/2$ but decays as $B(r)\propto\beta{u}(r)$, which suggests that the exact asymptotic limit of the direct correlation function $c(r)\to-\beta{u}(r)$ should also be violated. In fact, this can be rigorously proven from the asymptotics of the non-linear closure condition, see Eq.(\ref{eq:theory_oz_closure}). It is known that the asymptotic range provides large contributions to the above $\mu_{\mathrm{ex}}$ expression, which explains why the statistical route should be avoided. Naturally, this brings forth an inherent problem of the SHNC approach: its thermodynamic inconsistency.

\begin{figure}[t]
	\centering
	\includegraphics[width=3.40in]{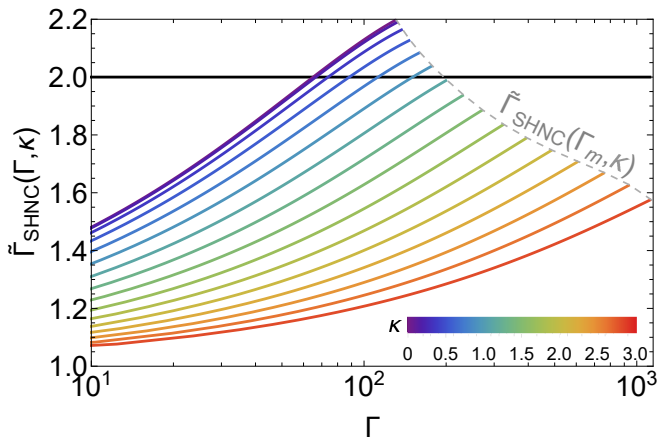}
\caption{Graphical representation of the normalized mapping of the SHNC approach, $\tilde{\Gamma}_{\mathrm{SHNC}}(\Gamma,\kappa)=\Gamma_{\mathrm{SHNC}}(\Gamma,\kappa)/\Gamma$, as a function of the coupling parameter $\Gamma$. Each color-coded curve corresponds to a different value of the screening parameter $\kappa$ within the range $\kappa\in[0,3]$. The gray dashed line demarcates the upper limit of validity of the SHNC approach obtained by evaluating $\tilde{\Gamma}_{\mathrm{SHNC}}(\Gamma,\kappa)$ at $\Gamma=\Gamma_{\mathrm{m}}(\kappa)$, where $\Gamma_{\mathrm{m}}(\kappa)$ is given by Eq.(\ref{eq:Gamma_melting}). The black horizontal line illustrates the simplified mapping $\tilde{\Gamma}_{\mathrm{SHNC}}(\Gamma,\kappa)=2$ that emerges by applying the T/2-HNC approach to the 2D-YOCP.}\label{fig:SHNC_scaling}
\end{figure}

\subsection{Comments on the T/2-HNC approach and the metastable states}\label{sec:iet_meta}

\noindent The T/2-HNC approach utilizes the HNC approximation at a reduced half temperature. For the YOCP, this approach corresponds to a simplified version of the SHNC approach for which $\Gamma_{\mathrm{SHNC}}(\Gamma,\kappa)=2\Gamma$. This empirical approach was recently applied to determine the glass transition line of the 2D-YOCP with the aid of mode coupling theory\,\cite{yazdi2015}. This was carried out without any discussion concerning the validity of the T/2-HNC approximation for metastable or even for stable 2D-YOCP liquids. Such an analysis will be performed in what follows.

The T/2-HNC approach was initially proposed for two-dimensional binary mixtures of point-dipoles interacting via a $\propto{r}^{-3}$ pair potential. For such systems it was observed empirically that, if the state point temperature is rescaled from $T$ to $T/2$, then the HNC approach can be employed for fairly accurate estimates of the structural properties\,\cite{hajnal2011}. Given the dependence of bridge functions on the softness (see the successes of the HNC and of the Percus-Yevick approaches) and the empirical nature of the re-scaling, it is evident that the T/2-HNC approach should not be applied to other systems without prior verification of its accuracy. Some work in this direction was performed in Ref.\cite{yazdi2015}, where some qualitative agreement between Monte Carlo simulations and the T/2-HNC approximation was reported for the 2D YOCP. However, (a) these simulations were performed in the stable fluid regime, thus the extension to the supercooled regime involves an unjustified extrapolation, (b) the documented accuracy of the SHNC approach within the stable fluid region and the strong deviations of the SHNC mapping from the T/2-HNC mapping (see Fig.\ref{fig:SHNC_scaling}) prove that the T/2-HNC approach does not lead to accurate 2D-YOCP structural properties even in the stable fluid region.

Generally speaking, any approximate integral equation theory closure that is derived from computer simulations should only be used within its range of validity (determined by the simulation input employed to construct it). Extrapolations outside the original range of validity are sometimes possible\,\cite{lucco2021b}, but should always be performed with great care. Hence, considering that the range of validity of the SHNC approach is $\Gamma/\Gamma_{\mathrm{m}}(\kappa)<1.0$ (see the dashed gray line in Fig.\ref{fig:SHNC_scaling}), that the T/2-HNC approach performs poorly even in the stable fluid region and that the HNC approach is expected to perform poorly in the supercooled regime\,\cite{lucco2021b}, it can be concluded that, at the moment, there is no integral equation theory approximation which can be employed to accurately predict the structural properties of supercooled 2D-YOCP liquids.

\section{Summary and future work}\label{sec:summary}

\noindent The structural and thermodynamic properties of dense two-dimensional Yukawa liquids were extensively investigated with molecular dynamics simulations. The "exact" thermodynamic properties were employed in order to construct a new equation of state for the excess internal energy valid in the parameter regime most relevant for contemporary experiments. Our equation of state exhibited excellent agreement with the simulation results and, contrary to most 2D YOCP equations of state available in the literature, proved to be robust with respect to thermodynamic integration and differentiation. The "exact" structural properties were employed to formulate the scaled hypernetted-chain approach that is constructed by up-scaling the interaction strength until the bare HNC recovers the exact magnitude of the first peak of the radial distribution function. The SHNC was demonstrated to significantly improve the HNC structural predictions and to achieve a $2\%$ accuracy in thermodynamic quantities.

For future improvement of the present results, it would be important to confirm that the 2D-YOCP is R-simple, to numerically trace out multiple isomorphic lines and to determine an accurate analytical parameterization of these isomorphs. This would allow for the construction of more accurate equations of state in the spirit of the Rosenfeld-Tarazona scaling and would also allow for the development of the isomorph-based empirically modified hypernetted chain approach for the 2D-YOCP. The latter integral equation theory approximation should lead to unprecedented levels of accuracy superior to that of the scaled hypernetted chain approach, but it requires an analytical parametrization for the 2D-OCP bridge function whose extraction from simulations is a formidable task.

\section*{Acknowledgments}

\noindent The authors would like to acknowledge the financial support of the Swedish National Space Agency under grant no.\,143/16. Molecular dynamics simulations were carried out on resources provided by the Swedish National Infrastructure for Computing (SNIC) at the NSC (Link{\"o}ping University) that is partially funded by the Swedish Research Council through grant agreement no.\,2018-05973.


\end{document}